# Interstitial diffusion of ion-implanted boron in crystalline silicon

**O.I. Velichko, A.P. Kavaliova**
Belarusian State University of Informatics and Radioelectronics
*velichkomail@gmail.com*

Modeling of the long-range migration of boron interstitials during low temperature annealing of ion-implanted silicon crystals has been carried out.

**ВВЕДЕНИЕ**

В настоящее время для формирования мелких ***p-n*** переходов широко используется высокодозное внедрение ионов с низкой энергией имплантации. В процессе имплантации и во время последующей термообработки ионно-имплантированных слоев генерируется большое количество неравновесных межузельных атомов кремния, что приводит к скоротечной диффузии атомов примеси и уширению профиля распределения примесных атомов. Поэтому, для формирования легированных областей с крутыми распределениями примеси в вертикальном и боковых направлениях необходимо применять отжиги с чрезвычайно малым термическим бюджетом [1]. Один из способов обеспечения малого термического бюджета является использование низкотемпературного отжига короткой продолжительности [2]. Также широко используется имплантация бора в слой кремния, предварительно аморфизованном внедрением ионов Ge [3]. В результате твердофазной эпитаксиальной рекристаллизация такого аморфного слоя образуется весьма совершенная структура без видимых посредством электронной микроскопии дефектов. Однако, скоротечная диффузия наблюдается и в рекристаллизованных слоях, хотя имеет совершенно иной характер. Действительно, при низких температурах отжига во всех упомянутых выше случаях легирования формируется протяженный "хвост" на профиле распределения бора в области низкой концентрации примесных атомов. Этот "хвост" представляет прямую линию при логарифмическом масштабе по оси концентрации. Как показано в [4,5], большинство случаев формирования "хвостов" в области низкой концентрации примесных атомов, включая имплантацию бора в предварительно аморфизованный кремний, связано с явлением длиннопробежной миграции неравновесных межузельных атомов примеси. В работе [6] подобное исследование было выполнено для имплантации бора в кристаллический кремний в диапазоне средних доз, недостаточных для создания аморфного слоя. Было найдено, что средняя длина пробега межузельных атомов бора $l_{AI}$ равна 24 нанометрам при температуре отжига 600 ºС и длительности 10 с. Это значение в два раза больше, чем средняя длина пробега межузельных атомов бора $l_{AI}$ = 12 нм и $l_{AI}$ = 11 нм для перераспределения бора в предварительно аморфизованном и рекристаллизованном кремнии при отжиге продолжительностью 60 с и температурах 850 ºС [4] и 800 ºС [7] соответственно. Следует отметить, что из-за более высоких температур и большей длительности отжига в двух последних случаях имело место существенное увеличение термического бюджета. Поэтому, имеет смысл исследовать механизм формирования "хвоста" для случая имплантации бора в кристаллический кремний и последующего низкотемпературного отжига с длительностью большей 10 секунд.

## 1. МОДЕЛЬ МЕЖУЗЕЛЬНОЙ ДИФФУЗИИ БОРА

Для моделирования межузельной диффузии атомов бора используем модель, предложенную в [4,5], которая включает следующую систему уравнений:

**1. Закон сохранения для неподвижных атомов примеси:**

$$\frac{\partial C^T(x,t)}{\partial t} = \frac{C^{AI}(x,t)}{\tau^{AI}} - G^{AI}(x,t) , \qquad (1)$$

**2. Уравнение диффузии неравновесных межузельных атомов бора:**

$$d^{AI} \frac{\partial^2 C^{AI}}{\partial x^2} - \frac{C^{AI}}{\tau^{AI}} + G^{AI}(x,t) = 0 . \qquad (2)$$

Здесь $C^T$ — полная концентрация атомов примеси в положении замещения, связанных в кластеры и захваченных протяженными дефектами (неподвижные атомы бора); $C^{AI}$ — суммарная концентрация неравновесных межузельных атомов бора в различных зарядовых состояниях; $d^{AI}$ и $\tau^{AI}$ — коэффициент диффузии и среднее время жизни неравновесных межузельных атомов примеси, соответственно; $G^{AI}$ — скорость генерации межузельных атомов примеси. Концентрация мобильных межузельных атомов примеси также включена в $C^T$.

В отличие от модели [4,5], было использовано распределение Пирсон-IV для описания пространственного распределения атомов бора после имплантации. С другой стороны, для упрощения расчета диффузии примеси пространственное распределение скорости генерации неравновесных межузельных атомов бора описывалось распределением Гаусса, максимум которого был сдвинут на место рассчитанного ранее положения максимума концентрации имплантированной примеси. Также учитывалась межузельная диффузия небольшой части атомов бора в процессе имплантации ионов [6].

## 2. РЕЗУЛЬТАТЫ МОДЕЛИРОВАНИЯ

Результаты моделирования межузельной диффузии атомов бора в рамках описанной выше модели представлены на Рис. 1 и Рис. 2. Для сравнения использовались экспериментальные данные [2]. В работе [2] имплантация ионов бора была выполнена с дозой 1×10$^{14}$ см$^{-2}$ и энергией 1 кэВ в кремниевые подложки ориентации (100) в условиях, препятствующих каналированию ионов. Перед внедрением ионов был эпитаксиально выращен слой кремния толщиной 10 мкм с проводимость *p*-типа и удельным сопротивлением 10 Ом·см. Профили распределения концентрации атомов бора перед началом и после термообработки были измерены методом вторичной ионной масс-спектроскопии с высоким разрешением по глубине. Отжиг образцов был проведен при температуре 600 ºC с длительностью 10, 600 и 1200 с.

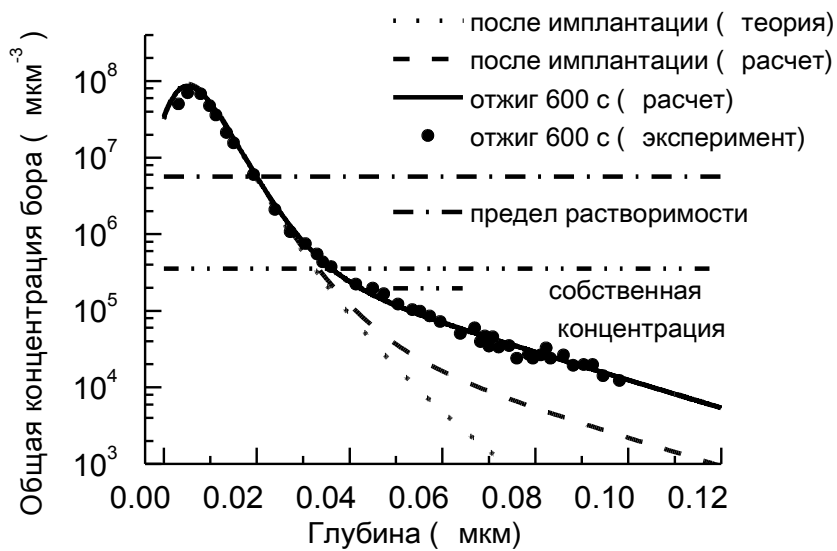

*Рис.1*. Рассчитанный профиль распределения концентрации бора после отжига в течение 600 с при температуре 600 °C. Пунктирная кривая представляет профиль распределения концентрации примеси после имплантации согласно распределению Пирсон-IV. Штриховая кривая — расчет начального распределения в предположении длиннопробежной миграция части атомов бора во время внедрения ионов, который хорошо согласуется с экспериментом. Экспериментальные данные по распределению бора после отжига (зачерненные кружки) взяты из [2]

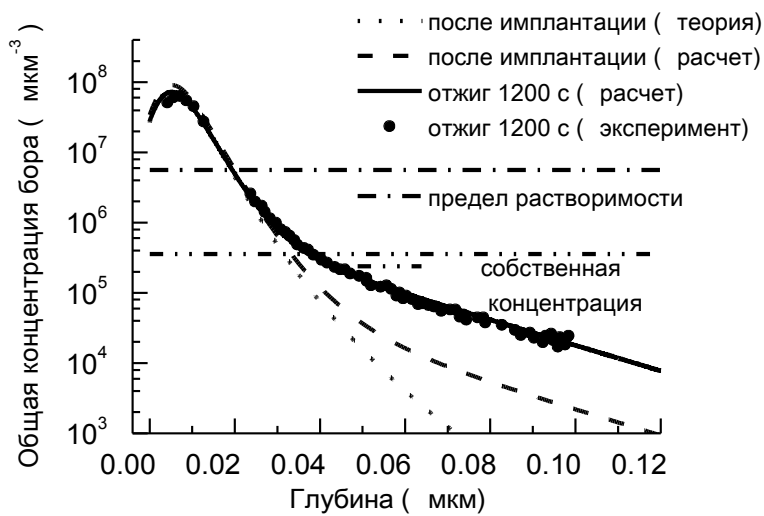

*Рис.2*. Рассчитанный профиль распределения концентрации бора после отжига в течение 1200 с при температуре 600 °C. Экспериментальные данные (зачерненные кружки) взяты из [2]

Как видно из Рис. 1 и Рис. 2, результаты моделирования перераспределения бора в течение отжига длительностью 600 и 1200 с хорошо согласуются с экспериментальными данными, если используется то же самое значение средней длины миграции межузельных атомов примеси $l_{AI} = \sqrt{d^{AI}\tau^{AI}} = 24$ нм, которое было получено в [6] для длительности отжига 10 с. Отсюда следует, что длина пробега межузельных атомов бора практически не меняется с увеличением длительности термообработки и совпадает со средней длиной миграции межузельных атомов примеси во время имплантации ионов [6]. Это подтверждает корректность разработанной модели длиннопробежной миграции неравновесных межузельных атомов бора в ионно-имплантированных слоях кремния в процессе низкотемпературного отжига.

Наилучшее совпадение с экспериментальными профилями распределения концентрации бора после отжига, представленными на Рис. 1 и Рис. 2, было получено в предположении, что соответственно 1.5 % and 2.29 % внедренных атомов бора участвовали в скоротечной межузельной диффузии и затем опять стали неподвижными, перейдя в положение замещения или образовав комплексы с дефектами кристаллической структуры. Следует отметить, что полученное значение $l_{AI} = 24$ нм для разных длительностей обработки приблизительно в два раза больше, чем средняя длина пробега межузельных атомов бора в слоях кремния, предварительно аморфизованных имплантацией ионов Ge и рекристаллизованных на начальной стадии отжига. Это означает, что мигрирующие межузельные атомы бора интенсивно взаимодействуют с атомами Ge, которые использовались для формирования приповерхностного аморфного слоя перед внедрением ионов бора.

## ЛИТЕРАТУРА